\documentclass[amsmath,amssymb,aps,prc,reprint,superscriptaddress]{revtex4-2}
\usepackage{graphicx}
\usepackage{dcolumn}
\usepackage{bm}
\usepackage{physics, hyperref}
\usepackage{color}

\renewcommand{\vec}[1]{\mbox{\boldmath $#1$}}
\usepackage[normalem]{ulem}
\usepackage{xcolor}
\renewcommand\sout{\bgroup\markoverwith{%
  \textcolor[rgb]{1,0.0,0.0}{%
    \rule[.35ex]{2pt}{0.6pt}%
    \hspace{-2pt}%
    \rule[.65ex]{2pt}{0.6pt}%
  }%
}\ULon}

\begin{document}

\title{Microscopic description of $^{12}$C+$^{12,13}$C fusion reactions at nuclear astrophysical 
energies }

\author{K. Nagao}
\affiliation{Department of Physics, Kyoto University, Kyoto 606-8502, Japan}
\author{K. Hagino}
\affiliation{Department of Physics, Kyoto University, Kyoto 606-8502, Japan}
\affiliation{Institute for Liberal Arts and Sciences, Kyoto University, Kyoto 606-8501, Japan}
\affiliation{ 
RIKEN Nishina Center for Accelerator-based Science, RIKEN, Wako 351-0198, Japan
}
\author{K. Uzawa}
\affiliation{Nuclear Data Center, Japan Atomic Energy Agency, Tokai 319-1195, Japan}
\affiliation{ 
RIKEN Nishina Center for Accelerator-based Science, RIKEN, Wako 351-0198, Japan
}
\date{\today}

\begin{abstract}
The $^{12}$C + $^{12}$C 
fusion reaction plays a key role in several astrophysical phenomena. However, 
it is difficult to determine 
its cross sections in the relevant energy region 
because of both 
low cross sections 
and strong resonant structures. 
On the other hand, 
the $^{12}$C + $^{13}$C 
system 
shows a much smoother energy dependence of fusion cross sections. 
To simultaneously analyze the $^{12}$C + $^{12,13}$C systems,  
we here develop a reaction model that explicitly treats the entrance channel and the compound nucleus states. 
For this purpose, we combine the discrete basis model for the entrance channel 
and the shell model for the compound nuclei. 
The coupling strengths between the entrance channel and the compound nucleus states 
are determined so that the 
fusion cross sections for these systems match with each other at the resonance energies for the $^{12}$C + $^{12}$C system, 
as has been observed experimentally. 
The model successfully reproduces the significantly different behaviors of fusion 
cross sections 
in these systems. 
\end{abstract}
\maketitle

\section{Introduction\label{sec:introduction}}
The $^{12}\mathrm{C} + {}^{12}\mathrm{C}$ fusion reaction plays an important role in astrophysical phenomena such as the evolution of massive stars~\cite{10.1111/j.1365-2966.2012.20193.x}, X-ray superbursts~\cite{Cooper_2009} and Type Ia supernovae~\cite{Bravo2011}. 
However, a direct measurement of the fusion cross sections at astrophysical energies is extremely difficult because of the Coulomb barrier. 
Therefore, the fusion cross sections at 
these energies
need to be estimated by extrapolating experimental data obtained at higher energies.

Experimentally, the $^{12}\mathrm{C} + {}^{12}\mathrm{C}$ fusion reaction has been studied for many years~\cite{PhysRevLett.4.515, PhysRevLett.124.192701,PhysRevLett.124.192702, PhysRevC.73.064601,HIGH1977181, PhysRevLett.98.122501, PhysRevC.110.035808, PhysRevC.97.012801}. 
It is well known that the fusion excitation function for $^{12}\mathrm{C}+{}^{12}\mathrm{C}$ 
exhibits pronounced resonance structures at sub-barrier energies.
These resonance structures lead to a large uncertainty in the extrapolation procedure towards lower energies. 
Recently direct measurements have been successfully carried out using the charged-particle and $\gamma$-ray coincidence techniques~\cite{PhysRevLett.124.192701,PhysRevLett.124.192702, PhysRevC.97.012801,PhysRevC.110.035808}.
Nevertheless, significant uncertainties still 
remain around $E_\text{c.m.} \sim 2.2\, \mathrm{MeV}$, that is close to the Gamow energy 
for the carbon burning phase in massive stars.
The indirect measurement, that is, 
the Trojan Horse Method (THM), has also been applied to the $^{12}\mathrm{C} + {}^{12}\mathrm{C}$ fusion reaction~\cite{Tumino2018}. 
In this approach, the $^{12}\mathrm{C}+{}^{12}\mathrm{C}$ fusion cross section was extracted from measurements of the $^{14}\mathrm{N} + {}^{12}\mathrm{C}$ reaction, where the deuteron in $^{14}\mathrm{N}$ was treated as a quasi-free spectator. The results show a sequence of 
pronounced resonance structures in fusion cross sections, with an increasing trend of cross sections 
as the incident energy is lowered. 
The effect of the Coulomb distortion on the THM method was also discussed in
Refs.~\cite{Beck2020,Mukhamedzanov2022}. 

In contrast to the 
$^{12}\mathrm{C}+{}^{12}\mathrm{C}$ system, 
fusion cross sections of the $^{12}\mathrm{C} + {}^{13}\mathrm{C}$ system show a 
smooth energy dependence with no pronounced resonance structure~\cite{PhysRevC.85.014607, ZHANG2020135170}. 
Ref. \cite{PhysRevLett.110.072701} argued that this contrast mainly reflects 
the compound-nucleus properties: that is, 
(i) the smaller fusion $Q$ value in $^{12}\mathrm{C} + {}^{12}\mathrm{C}$ leads to lower excitation energies in the compound nucleus $^{24}\mathrm{Mg}$ at the same incident energy, 
(ii) the level density of the even-even nucleus $^{24}\mathrm{Mg}$ is lower than that of $^{25}\mathrm{Mg}$ at the same excitation energy, 
and (iii) the identical spin-zero nuclei in the entrance channel restrict populated compound nucleus states to those with even 
spin and positive parity. 
As a result, the effective level density for the $^{12}\mathrm{C} + {}^{12}\mathrm{C}$ fusion 
is reduced by roughly an order of magnitude. 
This leads to the fact that the resonances in $^{24}$Mg tend to be isolated, with the level spacing $D$ comparable to or larger than the decay width $\Gamma$, 
whereas in $^{12}\mathrm{C} + {}^{13}\mathrm{C}$ the resonances overlap with each other, 
yielding a smooth fusion excitation function.
In addition, it has also been observed experimentally 
that the $^{12}\mathrm{C} + {}^{13}\mathrm{C}$ fusion cross sections provide upper limits for those of $^{12}\mathrm{C} + {}^{12}\mathrm{C}$ in the measured energy range~\cite{PhysRevC.85.014607,ZHANG2020135170}.

Theoretically, 
various approaches have been applied to the ${}^{12}\mathrm{C}+{}^{12}\mathrm{C}$ fusion reaction, including the coupled-channels approach~\cite{Esbensen2011,ASSUNCAO2013355,IMANISHI1968267, IMANISHI196933, 10.1143/PTP.59.465,FINK1972259}, 
the time-dependent wave packet approach~\cite{PhysRevC.97.055802} and
time-dependent Hartree-Fock based approaches~\cite{PhysRevC.102.061602,HEENEN1981298,PhysRevC.100.024619}.
In these calculations, a phenomenological imaginary potential or the incoming boundary condition, which corresponds to a strong short-range absorption, 
are used to describe the formation of the compound nucleus.
Since the optical potential arises through the projection on a restricted model space\cite{FESHBACH1958357,FESHBACH1962287}, it should in principle contain 
information on the eliminated compound states.
However, it is almost impossible to take into account all the relevant channels, and thus the 
phenomenological imaginary potentials account for only the average effect 
of compound-nucleus resonances.

As a more microscopic calculation, Taniguchi and Kimura recently performed calculations for the ${}^{12}\mathrm{C}+{}^{12}\mathrm{C}$ fusion cross sections using the anti-symmetrized molecular dynamics (AMD) method~\cite{TANIGUCHI2021136790,TANIGUCHI2024138434}. 
Their work provided parameter-free calculations of the $^{12}\mathrm{C} + {}^{12}\mathrm{C}$ fusion cross sections, successfully predicting the emergence of resonance structures at low energies.
More recently, using the multichannel resonating group method (RGM), Descouvemont presented a fully microscopic description of the ${}^{12}\mathrm{C}+{}^{12}\mathrm{C}$ fusion reaction, explicitly incorporating the ${}^{12}\mathrm{C}+{}^{12}\mathrm{C}$ and $\alpha+{}^{20}\mathrm{Ne}$ channels
~\cite{tw8t-gsxg}. 
This calculation, allowing also excitations of the fragments, predicted resonance structures in the energy region beyond the range covered by the existing experimental data.
Although these studies have provided important insights into the fusion mechanism, it remains difficult to treat the $^{12}\mathrm{C}+{}^{12}\mathrm{C}$ and the $^{12}\mathrm{C}+{}^{13}\mathrm{C}$ systems within a common microscopic framework. In approaches such as AMD and RGM, the $^{12}\mathrm{C}+{}^{13}\mathrm{C}$ system would require a significantly larger model space, making a unified treatment of the two systems difficult.

As far as we know, 
there has been no attempt to simultaneously reproduce the excitation functions for these two $^{12}\mathrm{C} + {}^{12,13}\mathrm{C}$ systems within a unified framework that explicitly incorporates the compound nucleus states. 
Given this situation, the aim of this paper is to 
construct a reaction model that couples the $\mathrm{C}+\mathrm{C}$ channel to the compound nucleus states of the Mg nuclei.
With this model, we attempt to 
(i) provide a theoretical description of the $^{12}\mathrm{C} + {}^{12}\mathrm{C}$ fusion cross sections in the low-energy region relevant to nuclear astrophysics, and (ii) 
reproduce the different behaviors in fusion cross sections between the 
$^{12}$C+$^{12}$C and the $^{12}$C+$^{13}$C systems. 

Our approach is an extension of Refs.~\cite{PhysRevC.98.014604,s63h-y87h}. 
In Ref. \cite{PhysRevC.98.014604}, compound nucleus reactions with neutron incident in 
the $s$-wave 
were modeled by coupling the reaction coordinate to compound nucleus states described 
with a random matrix. 
In Ref. \cite{s63h-y87h}, a schematic one-dimensional model was developed to 
understand a resonance structure in transmission coefficients due to the 
isolated resonances.  
In this work, we extend these approaches by 
considering the charged-particle systems 
as well as non-zero partial waves.
Moreover, we employ a more microscopic treatment for the compound nuclei 
using realistic shell-model calculations. At the same time, we estimate the decay widths of the compound 
nucleus states using the statistical model. 
From this viewpoint, the present framework can be regarded as 
a microscopic extension 
of the phenomenological optical-potential approach
by explicitly treating 
the compound-nucleus contribution on the basis of microscopic structure information.

The paper is organized as follows. In Sec.~\ref{sec:hamiltonian} we detail the Hamiltonian which 
we employ in this paper. In Sec.~\ref{sec:sa_and_sc}, we present the detailed procedure for 
calculations of the fusion cross sections. In Sec.~\ref{sec:result}, we 
set up the model Hamiltonians 
and present the calculated results. 
Finally, we summarize the paper in Sec.~\ref{sec:summary}.

\section{Model Hamiltonian\label{sec:hamiltonian}}

\begin{figure}
    \centering
    \includegraphics[width=1.0\linewidth]{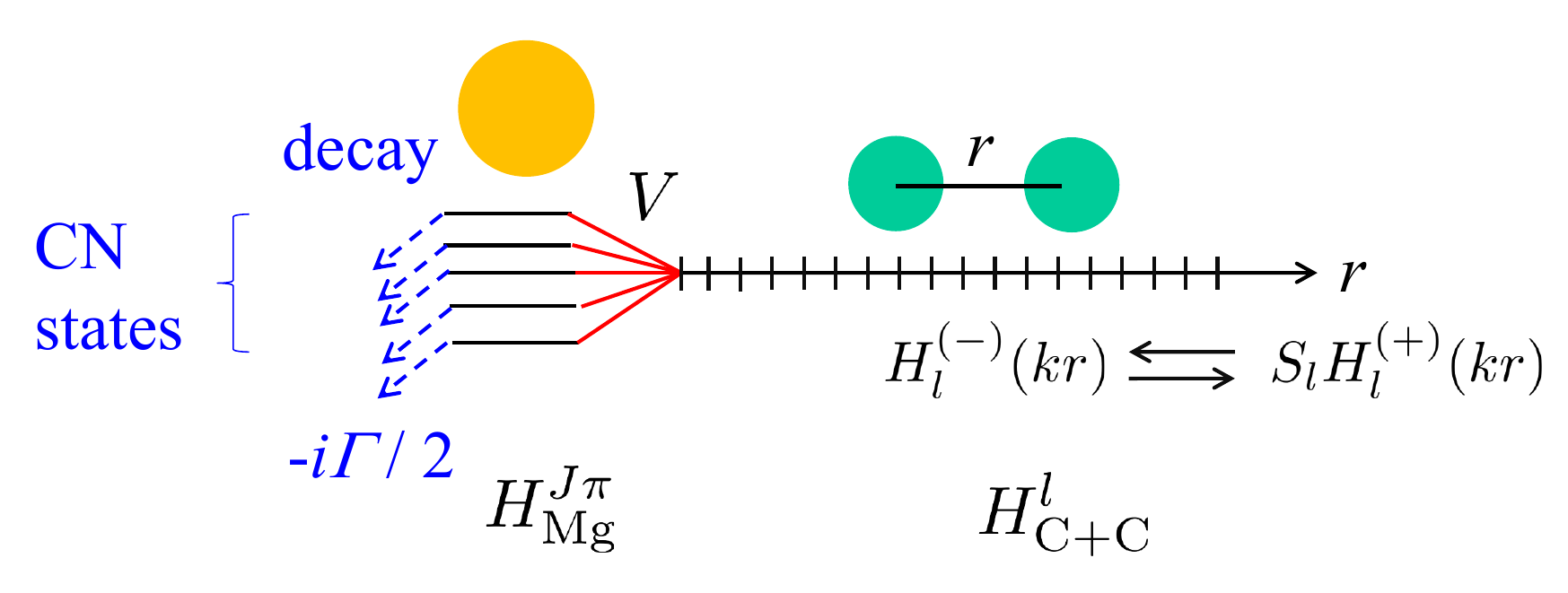}
    \caption{
    A schematic illustration of the model Hamiltonian given by Eq.~\eqref{eq:total_hamiltonian}. 
    The entrance C + C channel is treated with a standard potential model in the mesh representation. The compound nucleus states of Mg 
    consist of eigenvalues of a shell model Hamiltonian with decay widths $\Gamma$ estimated with a statistical model. The coupling between the C + C channel and the
    compound nucleus states of Mg is described by the matrix $V$. 
    }
    \label{fig:schematic_figure}
\end{figure}

The model used in this paper is based on the same form as that employed in Refs.\cite{PhysRevC.98.014604, s63h-y87h}. 
This model allows one to treat simultaneously 
the scattering states ($\mathrm{C} + \mathrm{C}$) and the 
compound nucleus states ($\mathrm{Mg}$).
The $\mathrm{C} + \mathrm{C}$ channel is described by the mesh representation, that is, the 
discrete basis method \cite{PhysRevC.107.044615,PhysRevC.98.014604,BERTSCH201968,ALHASSID2020168233,ALHASSID2021168381,PhysRevC.109.034611,PhysRevC.109.054606,s63h-y87h}, while 
the 
states of Mg are
calculated with a shell model.
The total Hamiltonian, specified by the relative orbital angular momentum $l$ in the $\mathrm{C} + \mathrm{C}$ channel and the spin-parity $J^\pi$ of the compound nucleus $\mathrm{Mg}$, 
is given 
in a matrix form as
\begin{eqnarray}
  \bm{H}^{l, J^\pi} = \begin{pmatrix}
    \bm{H}^{l}_{\text{C+C}} & \bm{V} \\
    \bm{V}^T & \bm{H}^{J^\pi}_{\text{Mg}}
  \end{pmatrix}\label{eq:total_hamiltonian},
\end{eqnarray}
where $\bm{H}^{l}_{\text{C+C}}$ is the Hamiltonian for the $\mathrm{C} + \mathrm{C}$ channel with the relative orbital angular momentum $l$, $\bm{H}^{J^\pi}_{\text{Mg}}$ is the Hamiltonian for the compound nucleus states with spin-parity $J^\pi$, and $\bm{V}$ is the coupling matrix. 
This Hamiltonian is schematically illustrated in Fig.~\ref{fig:schematic_figure}. 
Notice that a similar Hamiltonian has been used for ab-initio calculations 
of nucleon-nucleus scattering based on the no-core shell model \cite{PhysRevC.79.044606}.

Each component of the Hamiltonian (\ref{eq:total_hamiltonian}) is detailed in the next subsections.

\subsection{The C + C channel}

The Hamiltonian for the C+C channel, $\bm{H}^{l}_{\text{C+C}}$, describes scattering of two carbon 
nuclei. In this paper, for simplicity, we assume inert carbon nuclei, and neglect their excitations. Using the discrete basis 
formalism, the Hamiltonian matrix reads,
\begin{eqnarray}
  (\bm{H}^{l}_{\text{C+C}})_{ij} = \qty[2t + V_l(r_i)] \delta_{ij} - t \delta_{i, j+1} - t\delta_{i, j-1}\label{eq:c_c_hamiltonian},
\end{eqnarray}
where $i,j = (1, \dots, N_{\text{C+C}} )$ with 
$r_i = i \Delta r$,  $\Delta r$ being the mesh spacing. 
Here $N_{\text{C+C}}$ is the number of sites on the radial mesh, 
$t$ is given by $\hbar^2 / (2\mu \Delta r^2)$ with $\mu$ being reduced mass. 
$V_l(r_i)$ is a nucleus-nucleus potential, which is the sum of the Coulomb, the nuclear and 
the centrifugal potentials. 
For the nuclear part, we employ the M3Y + repulsion potential \cite{PhysRevC.75.034606}, 
which consists of a double-folding potential with the M3Y interaction and 
a similar double-folding potential with a repulsive interaction.  
The former is defined as
\begin{eqnarray}
    U_{\mathrm{M3Y}}(\bm{r}) = \int \dd{\bm{r}_1} \dd{\bm{r}_2} \rho_1(\bm{r}_1)\rho_2(\bm{r}_2) v_{\mathrm{M3Y}} (\bm{r} + \bm{r}_2 - \bm{r}_1 ),
    \label{eq:M3Y}
\end{eqnarray}
where $v_{\mathrm{M3Y}}$ is the M3Y effective nucleon-nucleon interaction~\cite{SATCHLER1979183}.
Here, we use the two parameter Fermi distribution function for the 
density distributions of the carbon nuclei:
\begin{align}
    \rho_i(\bm{r}) = \frac{\rho_0}{1 + \exp((r-R_{0i})/a_i)},
    \label{eq:density}
\end{align}
where $i (= 1, 2)$ labels each of the two nuclei.
For the repulsive part, 
a similar double folding procedure to Eq. (\ref{eq:M3Y}) is used, but with 
a zero-range interaction given by $v_r \delta(\bm{r})$. 
The density distributions for the repulsive part are 
the same as Eq. (\ref{eq:density}), except for the 
diffuseness parameter $a_r$, which is determined by fitting to the experimental 
fusion cross sections. 
With a given value of $a_r$, 
we follow Ref. \cite{PhysRevC.75.034606} and adjust 
the strength parameter $v_r$ so to reproduce the nuclear incompressibility of 
$K = 234\,\mathrm{MeV}$. 

In this way, 
there are three adjustable parameters in this potential, that is, $a$ and $R_0$ for the 
density distributions to be used in the M3Y part, and $a_r$ for the repulsive part. 
We take the same values of $a$ and $R_0$ as those in Ref.\cite{Esbensen2011}. 
On the other hand, we re-adjust the diffuseness parameter $a_r$ for the 
repulsive term, since we use a different reaction model from that in Ref. \cite{Esbensen2011}.

\begin{figure*}[tb]
  \centering
    \includegraphics[width=0.45\linewidth]{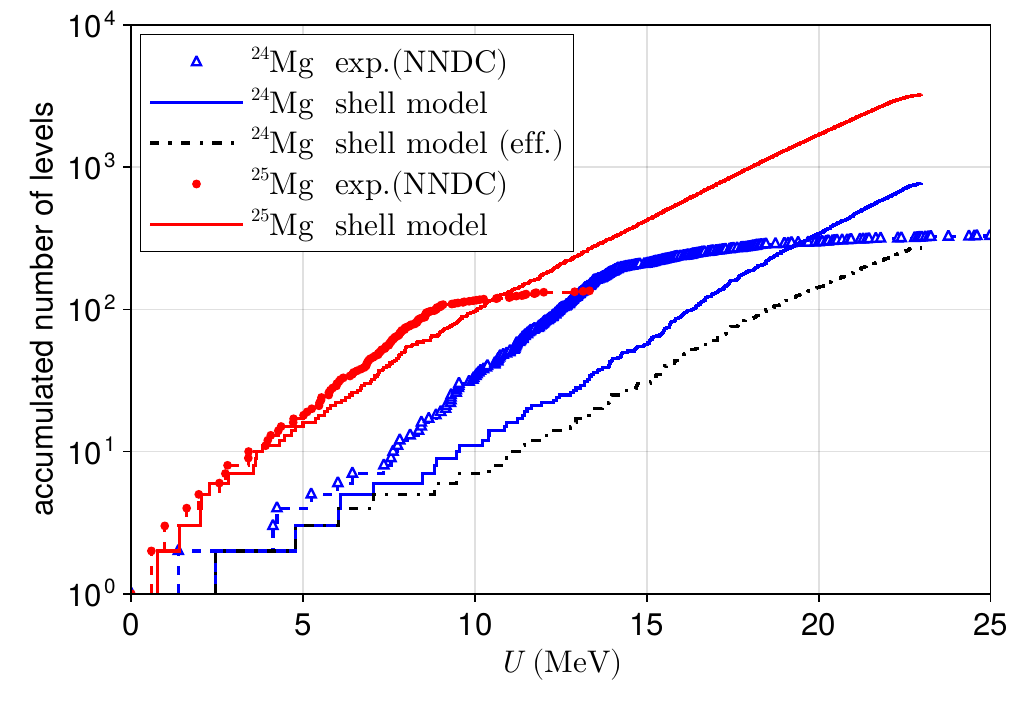}
    \includegraphics[width=0.45\linewidth]{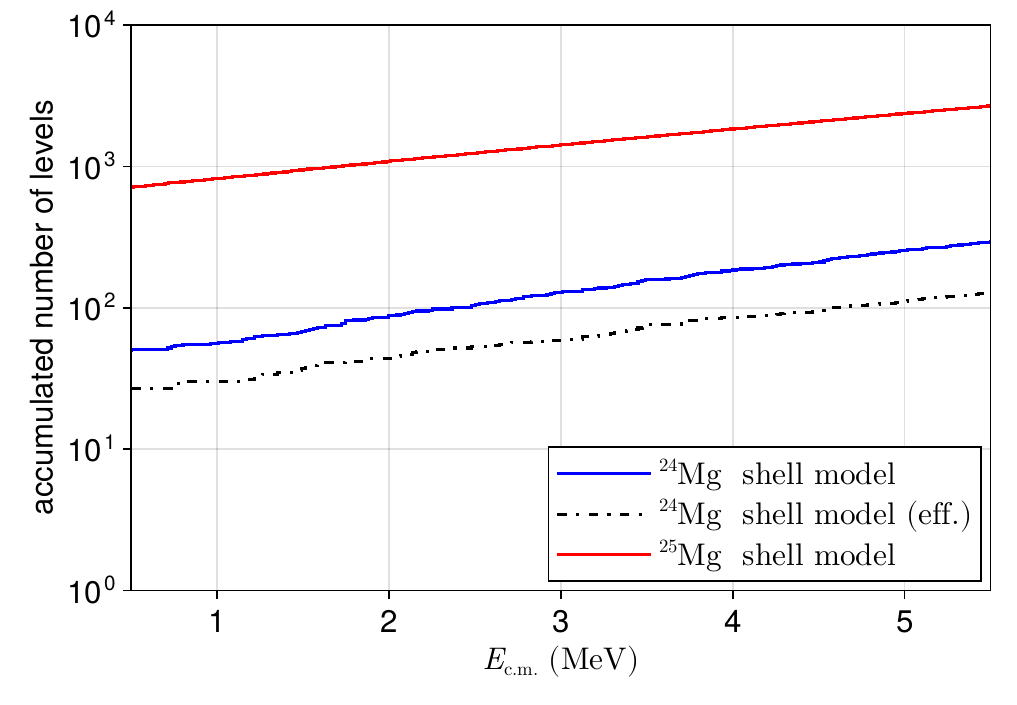}
  \caption{
  The accumulated number of levels of the $^{24,25}$Mg nuclei calculated with the shell model 
  code {\tt KSHELL} as a function of the excitation energy (the left panel) and of the center-of-mass 
  energy with Eq. \eqref{eq:q_value_and_e_cm} (the right panel). The results of $^{24}\mathrm{Mg}$ are shown by the blue lines, while the results of $^{25}\mathrm{Mg}$ are shown by the red lines. 
  The black dash-dotted lines show the effective accumulated number of levels for $^{24}\mathrm{Mg}$ with even spin and positive parity. For $^{25}\mathrm{Mg}$, levels are included up to $J=11/2$ for positive parity and up to $J = 9/2$ for negative parity. The filled circles and open triangles connected by dashed lines represent the experimental data of $^{25}\mathrm{Mg}$ and $^{24}\mathrm{Mg}$, respectively, taken from NNDC~\cite{NNDC_NuDat3}.  
  }
  \label{fig:accumulated_num_levels}
\end{figure*}

\subsection{The compound nucleus states of Mg}

The Hamiltonian $\bm{H}_{\text{Mg}}$
describes the energy and the decay width of the compound nucleus states. 
It is a diagonal matrix and reads, 
\begin{eqnarray}
  (\bm{H}^{J^\pi}_{\text{Mg}})_{\mu\nu} = \qty(\mathcal{E}^{J^\pi}_\mu - i \Gamma^{J^\pi}_\mu / 2) \delta_{\mu\nu}\label{eq:mg_hamiltonian},
\end{eqnarray}
where $\mathcal{E}_\mu$ is the energy of a compound nucleus state with ($\mu = 1 , \dots, N_{\text{Mg}}$) , 
while $\Gamma^{J^\pi}_\mu$ is the decay width for the state $\mu$. 

In this paper, we generate $\mathcal{E}_\mu$ using a shell model calculation. 
To this end, we use the computer code  \verb+KSHELL+ \cite{SHIMIZU2019372} 
with the \verb+sdpf-mu+\cite{PhysRevC.86.051301} interaction. 
The accumulated number of levels for $^{24,25}$Mg is shown 
in Fig.\ref{fig:accumulated_num_levels}. 
The left panel is shown as a function of the excitation energy $U$, while the right panel 
is plotted as a function of the corresponding center-of-mass energy $E_{\rm c.m.}$. 
Using the fusion $Q$ values, the excitation energy $U$ of the compound nucleus is related to the center-of-mass energy $E_\text{c.m.}$ as 
\begin{eqnarray}
  U - Q = E_\text{c.m.}. 
  \label{eq:q_value_and_e_cm}
\end{eqnarray}
The fusion $Q$ values are 13.934 MeV for $^{12}\mathrm{C} + {}^{12}\mathrm{C}$ and 16.318 MeV for $^{12}\mathrm{C} + {}^{13}\mathrm{C}$.
In the figures, the blue and the red curves denote the results for $^{24}$Mg and $^{25}$Mg, respectively. These are compared to the experimental data (the open triangles for $^{24}$Mg 
and the filled circles for $^{25}$Mg) \cite{NNDC_NuDat3}. The dash-dotted lines denote 
the results for the levels of $^{24}$Mg with even $J$ and the positive parity only, 
which are relevant to the $^{12}$C+$^{12}$C fusion. 

Comparing the effective number of levels for $^{24}\mathrm{Mg}$ (the dash-dotted lines) with the number of levels for $^{25}\mathrm{Mg}$ (the red solid line) 
at the same center-of-mass energy, one can find that they differ by more than an order of magnitude. 
In the case of $^{25}\mathrm{Mg}$, the accumulated number of levels is relatively well reproduced
with the present shell model calculation, as compared to the case of $^{24}\mathrm{Mg}$. 
We will discuss the effect of this deviation in $^{24}\mathrm{Mg}$ in Sec.\ref{sec:result}.

\begin{figure*}[tb]
  \centering
  \begin{minipage}{0.49\linewidth}
      \centering
    \includegraphics[width=\linewidth]{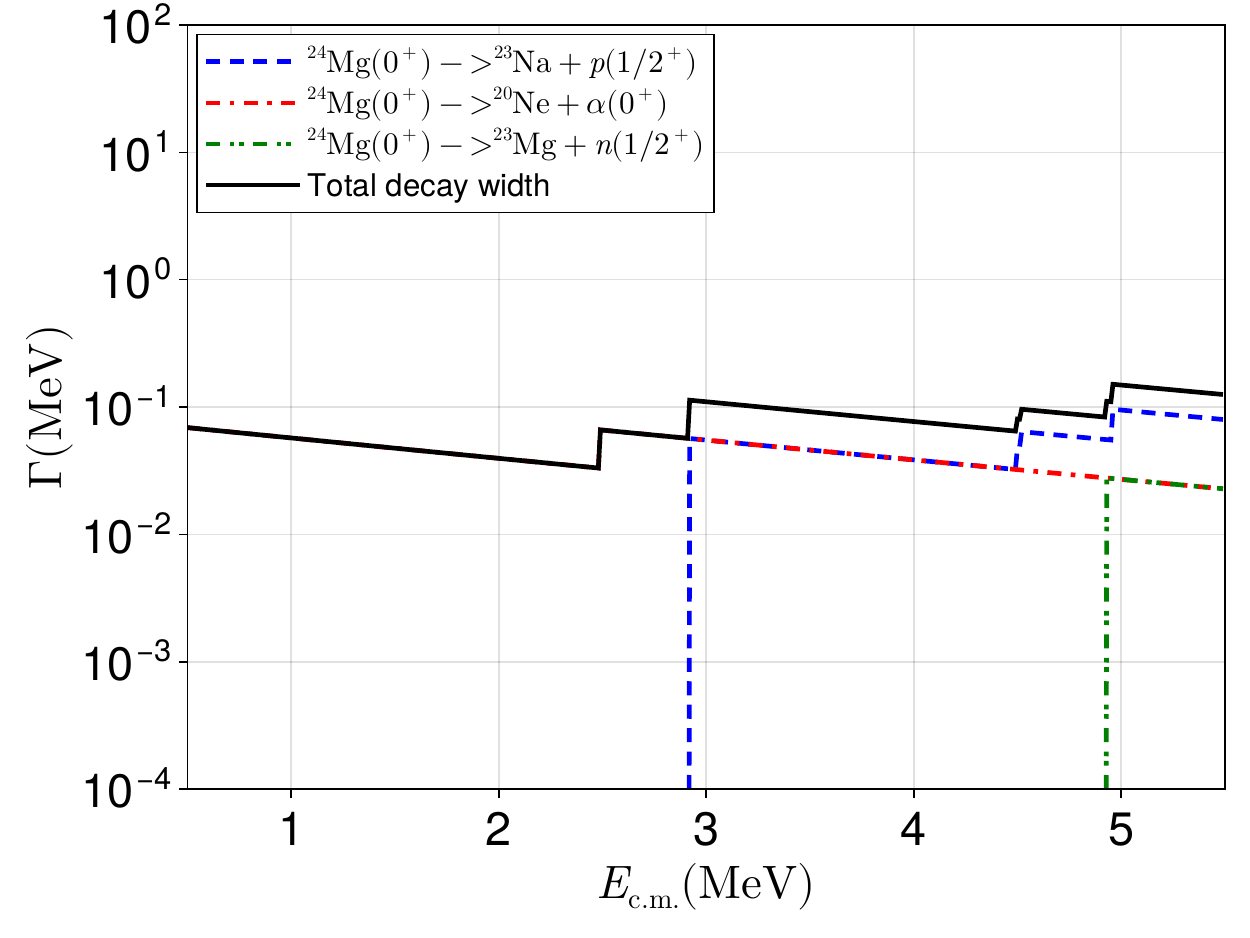}
  \end{minipage}
  \hfill
  \begin{minipage}{0.49\linewidth}
      \centering
    \includegraphics[width=\linewidth]{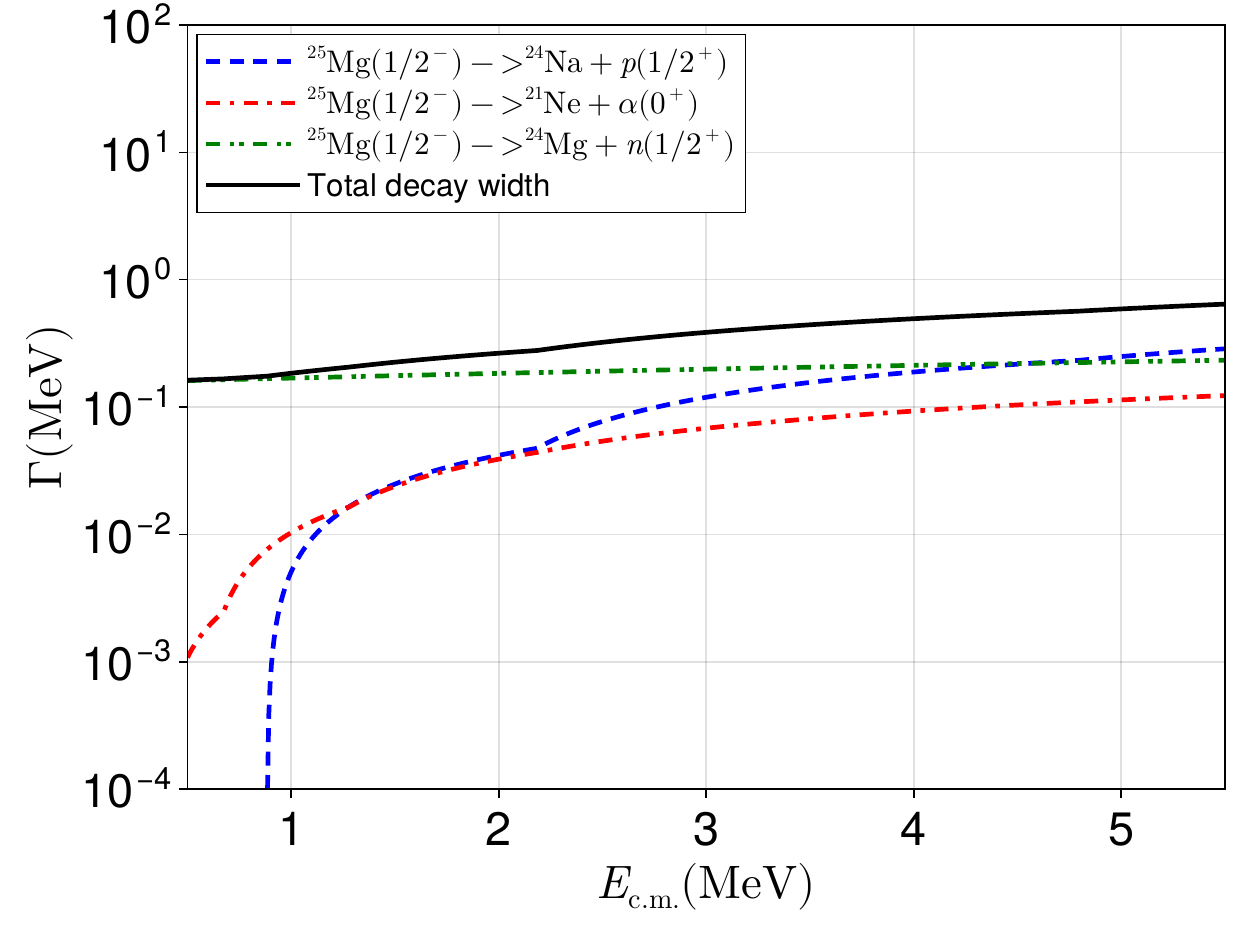}
  \end{minipage}
  \caption{
    Left panel: the decay width of $^{24}\mathrm{Mg}(J^\pi = 0^+)$ obtained by explicitly counting the number of final states $N(I_B)$. Right panel: the decay width of $^{25}\mathrm{Mg}(J^\pi = 1/2^-)$ obtained using the statistical model (Eq.~\eqref{eq:decay_width_statistical}).
    The blue dashed, the green dash-dot-dotted, and the red dash-dotted lines are the widths 
    for the $p$ decay, the $n$ decay, and the $\alpha$ decay, respectively.  
    }
    \label{fig:decay_width}
\end{figure*}

The decay widths $\Gamma_\mu$ are estimated with a statistical model~\cite{fröbrich1996theory}.
For a process where a compound nucleus state $C$ with energy $E^*_C$ and spin $J$ emits a light particle $y$ with spin $I_y$ and energy in the interval $(\mathcal{E}_b, \mathcal{E}_b + \dd{\mathcal{E}}_b)$, leaving a residual nucleus $B$ with spin $I_B$, 
the decay width is given by, 
\begin{widetext}
  \begin{eqnarray}
  \Gamma^{J}_{\{C;J;E_C^*\}\to\{b;I_y\,I_B;\mathcal{E}_b\}}
  =
  \frac{1}{2\pi \rho(E_C^*;J)}
  \sum_{S_b=\left|I_y-I_B\right|}^{I_y+I_B}
  \sum_{l_b=\left|J-S_b\right|}^{J+S_b}
  \int_0^{E_C^*-B_b} \dd{\mathcal{E}}_b\,
  \rho_B(E_C^*-B_b-\mathcal{E}_b;I_B)\,
  T^{J}_{l_b S_b}({\mathcal{E}}_b).
  \end{eqnarray}
where $\rho_B$ and $\rho$ are the level densities of the residual nucleus and the compound nucleus, respectively. 
The summation over $l_b$ is restricted to satisfy the parity conservation condition.
If one takes the classical 
transmission coefficient $T^{J}_{l_b S_b}({\mathcal{E}}_b)=\theta({\mathcal{E}}_b-V_B)$, where 
$\theta$ is a step function and $V_B$ is the barrier height
of the potential between the emitted particle $y$ and the residual nucleus $B$, 
the decay width becomes
  \begin{eqnarray}
    \Gamma^{J}_{\{C;J;E_C^*\}\to\{b;I_y\,I_B;\mathcal{E}_b\}}
    =
    \frac{1}{2\pi \rho(E_C^*; J)}
    \sum_{S_b=\left|I_y-I_B\right|}^{I_y+I_B}
    \sum_{l_b=\left|J-S_b\right|}^{J+S_b}
    \int_{V_B}^{E_C^*-B_b} \dd{{\mathcal{E}}}_b\,
    \rho_B(E_C^*-B_b-\mathcal{E}_b; I_B). 
  \label{eq:decay_width_statistical}
     \end{eqnarray}
\end{widetext}
From Eq.~\eqref{eq:decay_width_statistical}, one can find that the final integral term corresponds to the number of states of the residual nucleus, $N(I_B)$. 
That is, 
  \begin{equation}
    \Gamma^{J}_{\{C;J;E_C^*\}\to\{b;I_y\,I_B;\mathcal{E}_b\}}
    =
    \frac{1}{2\pi \rho(E_C^*; J)}
    N(I_B), 
    \label{eq:decay_width}
     \end{equation}
with
  \begin{equation}
      N(I_B)=     
      \sum_{S_b}
    \sum_{l_b}
    \int_{V_B}^{E_C^*-B_b} \dd{{\mathcal{E}}}_b\,
    \rho_B(E_C^*-B_b-\mathcal{E}_b; I_B). 
    \label{eq:number_of_state}
     \end{equation}

Notice that the decay width can be calculated according to Eq. (\ref{eq:decay_width_statistical})
once the level densities $\rho_B$ and $\rho$ are given. 
For simplicity, we employ a semi-empirical formula for the level density given by~\cite{ILJINOV1992517},
\begin{eqnarray}
  \rho(U) = \frac{\sqrt{\pi} a^{-1/4}}{12} (U - \Delta)^{-5/4} \exp\qty[2\sqrt{a(U-\Delta)}], 
\end{eqnarray}
where $a$ is the level density parameter. 
Here, $\Delta$ is the 
pairing energy given by
\begin{eqnarray}
  \Delta = \chi \frac{12}{\sqrt{A}}, 
\end{eqnarray}
with 
\begin{equation}
  \chi =
  \begin{cases}
    2 & \text{for even-even nuclei}, \\
    1 & \text{for even-odd nuclei}, \\
    0 & \text{for odd-odd nuclei}.
  \end{cases}
\end{equation}
Taking the spin dependence into account, the level density for the spin and parity 
of $I^\pi$ is written as~\cite{ILJINOV1992517} 
\begin{eqnarray}
  \rho(U, I^\pi) = \frac{2I + 1}{4\sqrt{2\pi}\sigma^3} \exp\qty[-\frac{(I+1/2)^2}{2\sigma^2}]\, \rho(U).
  \label{eq:level_density}
\end{eqnarray}
Here, $\sigma$ is the spin cutoff parameter, and using the nuclear mass $M$ and the radius $R = 1.2 A^{1/3}$, it is given by ~\cite{ILJINOV1992517} 
\begin{eqnarray}
  \sigma^2 = \sqrt{\frac{U-\Delta}{a}} \frac{\frac{2}{5}MR^2}{\hbar^2}.
\end{eqnarray}
The level density parameter $a$ is determined with the semi-empirical formula given in Ref.~\cite{ILJINOV1992517} for each nucleus.

For the $^{12}\mathrm{C}+{}^{12}\mathrm{C}$ fusion reaction, 
the energy differences between the initial states and the final states are small for the 
evaporation processes, 
and it is difficult 
to describe the level density statistically for the residual nuclei. 
Therefore, instead of using the statistical level density for the residual nuclei in Eq. \eqref{eq:number_of_state}, we evaluate the numerator of Eq. \eqref{eq:decay_width} by referring to the energy spectra of $^{23}\mathrm{Mg}$, $^{20}\mathrm{Ne}$, and $^{23}\mathrm{Na}$ from NNDC~\cite{NNDC_NuDat3} and explicitly counting the number of possible final states, $N(I_B)$. 
On the other hand, the level density of the initial compound nucleus appearing in the 
denominator of Eq.~\eqref{eq:decay_width} is evaluated using the statistical formula given in Eq.~\eqref{eq:level_density}, rather than being extracted directly from the shell-model spectrum. 
Notice that this prescription provides a smooth level density used in the evaluation of the decay widths.
Fig.~\ref{fig:decay_width} shows the decay widths so obtained. 
The left and the right panels show the results for $^{24}$Mg with $J^\pi=0^+$ 
and $^{25}$Mg with $J^\pi=1/2^-$, respectively.  
Here, the horizontal axis has been converted into the center-of-mass energy using Eq.~\eqref{eq:q_value_and_e_cm}.
It is found that the decay widths of $^{24}\mathrm{Mg}$ are slightly 
smaller than those of $^{25}\mathrm{Mg}$.

\subsection{The coupling matrix}

The coupling matrix $\bm{V}$ 
describes the couplings between the entrance channel and the compound nucleus states. 
We assume that the coupling takes place at a single position, $i_e$. 
The coupling matrix is thus given by,  
\begin{eqnarray}
  \bm{V}_{i, \mu} = \delta_{i, i_e} v_0 \qty(\Delta r)^{-1/2},
  \label{eq:coupling_mat}
\end{eqnarray}
where $v_0$ is a coupling strength parameter.
We here follow Ref. \cite{PhysRevC.98.014604} and scale the coupling strength with 
$(\Delta r)^{-1/2}$ so that the coupling is effectively independent of the mesh size. 
In this paper, we fix $i_e$ at the minimum of the potential $V_l(r_i)$ for each angular 
momentum $l$.
We also assume a state-independent coupling strength $v_0$ 
between the $\mathrm{C} + \mathrm{C}$ channel and the compound nucleus states of $\mathrm{Mg}$.

\section{scattering amplitude and cross sections\label{sec:sa_and_sc}}

Fusion cross sections are obtained by solving the eigen-value equation 
\begin{eqnarray}
  \bm{H}^{l, J^\pi} \vec{\psi}= \begin{pmatrix}
    \bm{H}^{l}_{\text{C+C}} & \bm{V} \\
    \bm{V}^T & \bm{H}^{J^\pi}_{\text{Mg}}
  \end{pmatrix}
   \begin{pmatrix}
    \vec{u} \\
    \vec{\phi}
  \end{pmatrix}
=E_{\rm c.m.}
   \begin{pmatrix}
    \vec{u} \\
    \vec{\phi}
  \end{pmatrix}, 
  \label{eq:eigenequation}
\end{eqnarray}
where $\vec{u}$ and $\vec{\phi}$ are the vectors in the C+C channel and the compound nucleus states of Mg, respectively.  
We impose $u(0) = 0$ so that the wave function is regular at the origin, $r=0$.
The asymptotic behavior of $u(r)$ at large $r$ 
for a scattering solution with a real and positive wave number $k$ is given by  
\begin{eqnarray}
  u(r) \rightarrow A(k)\qty[H_l^{(-)}(kr) - S^{l,J^\pi}_{C+C}(k) H_l^{(+)}(kr)]\label{eq:asym_boundary},
\end{eqnarray}
where $S^{l,J^\pi}_{C+C}(k)$ is the $S$-matrix for elastic scattering and $A(k)$ is an overall normalization factor. 
Due to the decay width $\Gamma$ for the compound nucleus states (see Eq. (\ref{eq:mg_hamiltonian})), 
the absolute value of the $S$ matrix becomes less than unity, that is, 
$|S^{l,J^\pi}_{C+C}(k)|<1$. In Eq. (\ref{eq:asym_boundary}), 
$H_l^{(\pm)}(kr)$ denote the incoming($-$) and the outgoing(+) Coulomb wave functions, which are 
defined by
\begin{eqnarray}
  H_l^{(\pm)}(kr) = G_l(kr) \pm i F_l(kr),
\end{eqnarray}
where $F_l$ and $G_l$ are the regular and the irregular Coulomb functions, respectively.

In Eq. (\ref{eq:eigenequation}), 
one must take into account the non-vanishing component of the wave function, $u(N_{\text{C+C}}+1)$, which is 
not in the vector space for the $\mathrm{C}+\mathrm{C}$ channel. 
This can be done by modifying Eq. (\ref{eq:eigenequation}) to 
\begin{eqnarray}
  \qty(E_{\rm c.m.} - \bm{H}^{l, J^{\pi}}) \vec{\psi} = \vec{h}, \label{eq:schro_eq_for_scatt}
\end{eqnarray}
where $h(i) = - t u(N_{\text{C+C}}+1)\delta_{i, N_{\text{C+C}}}$ \cite{PhysRevC.109.034611,PhysRevC.109.054606}.
Multiplying ~Eq. \eqref{eq:schro_eq_for_scatt} from the left by the Green's function
\begin{eqnarray}
  \bm{G}^{l, J^\pi} (E_{\rm c.m.}) = \qty(E_{\rm c.m.} - \bm{H}^{l, J^{\pi}}) ^{-1},
\end{eqnarray}
one obtains
\begin{eqnarray}
  \frac{u(N_{C+C})}{u(N_{C+C}+1)} = -t \bm{G}^{l, J^\pi} (E_{\rm c.m.})_{N_{\text{C+C}}, N_{\text{C+C}}}. \label{eq:frac_from_shcro}
\end{eqnarray}
With Eq.~\eqref{eq:asym_boundary}, this ratio is related to the $S$-matrix as,
  \begin{align}
    &\frac{u(N_\text{C+C})}{u(N_\text{C+C}+1)} \notag \\
    &= \frac{H_l^-(k\Delta r N_\text{C+C}) - S^{l,J^\pi}_{\text{C+C}} H_l^+(k\Delta r N_\text{C+C})}{H_l^-(k\Delta r (N_\text{C+C}+1)) - S^{l,J^\pi}_{\text{C+C}} H_l^+ (k\Delta r (N_\text{C+C} + 1))},  \label{eq:frac_from_boundary}
  \end{align}
from which one obtains 
\begin{widetext}
  \begin{align}
    S^{l, J^\pi}_{\text{C+C}} = \frac{H_l^- (k\Delta r N_\text{C+C}) + t \vec{G}^{l, J^\pi}(E)_{N_\text{C+C}, N_\text{C+C}} H^-_l(k\Delta r (N_\text{C+C}+1))}{H_l^+(k \Delta r N_\text{C+C})+ t \vec{G}^{l, J^\pi}(E)_{N_\text{C+C}, N_\text{C+C}} H_l^+(k\Delta r (N_\text{C+C} + 1))}.
  \end{align}
\end{widetext}

For the $^{12}\mathrm{C} + {}^{12}\mathrm{C}$ reaction, 
one has to take into account the fact that this is a reaction with two 
identical bosons\cite{PhysRevC.91.044617,PhysRevC.96.064615}. 
Since $^{12}\mathrm{C}$ is a $0^+$ boson in its ground state, the fusion cross section becomes
\begin{align}\label{eq:fus_cs_12_12}
\sigma_{\text{fus},^{12}\mathrm{C} + ^{12}\mathrm{C}}
= \frac{2\pi}{k^2}\sum_{l:\mathrm{even}}(2l+1)\qty(1-\bigl|S^{J^\pi=l^+}_l\bigr|^2).
\end{align}
On the other hand, for the $^{12}\mathrm{C} + {}^{13}\mathrm{C}$ reaction, the ground state of $^{13}\mathrm{C}$ has spin-parity of $\frac12^{-}$, and thus the cross section reads 
\begin{align}\label{eq:fus_cs_12_13}
  \sigma_{\text{fus},^{12}\mathrm{C} + ^{13}\mathrm{C}}
&= \frac{\pi}{k^2}\sum_{l}\sum_{J =\abs{ l - 1/2}}^{l+1/2}\frac{2J+1}{2}\qty(1-\bigl|S^{l,J^\pi}_l\bigr|^2), 
\end{align}
for the compound nucleus $^{25}$Mg with the parity $\pi=(-1)^{l+1}$. 
Here, $(2J+1)/2$ is the spin statistical factor, which is divided by 2 because the entrance channel has two spin degenerated states.

The fusion cross sections are then converted into the astrophysical 
$S^*$-factor defined as \cite{PhysRevC.85.014607}
\begin{eqnarray}
  S^*(E_{\rm c.m.}) = E_{\rm c.m.} \sigma(E_{\rm c.m.}) \exp(\frac{87.21}{\sqrt{E_{\rm c.m.}}} + 0.46 E_{\rm c.m.}).
\end{eqnarray}
Even though the factor in the exponent, 87.21 MeV$^{1/2}$, is different between the 
$^{12}$C+$^{12}$C and $^{12}$C+$^{13}$C reactions in the original definition of the astrophysical 
$S$-factor, $S(E)=E\sigma(E)e^{2\pi\eta}$, where $\eta$ is the Sommerfeld parameter, 
we here follow Ref. \cite{PhysRevC.85.014607} to use the same value for the two systems in order to 
remove a trivial scaling effect. 

\section{Results and Discussion\label{sec:result}}

\subsection{Determination of the parameters $a_r$ and $v_0$\label{subsec:parameter_v0}}

Let us now numerically solve Eq. (\ref{eq:schro_eq_for_scatt}) and discuss the energy dependence 
of the fusion cross sections for the 
$^{12}$C+$^{12,13}$C systems. 
To this end, we have to determine 
the diffuseness parameter $a_r$ in the repulsive term 
of the nuclear potential 
and the coupling strength parameter $v_0$ 
by fitting to the experimental data.

\begin{figure}
      \includegraphics[width=\linewidth]{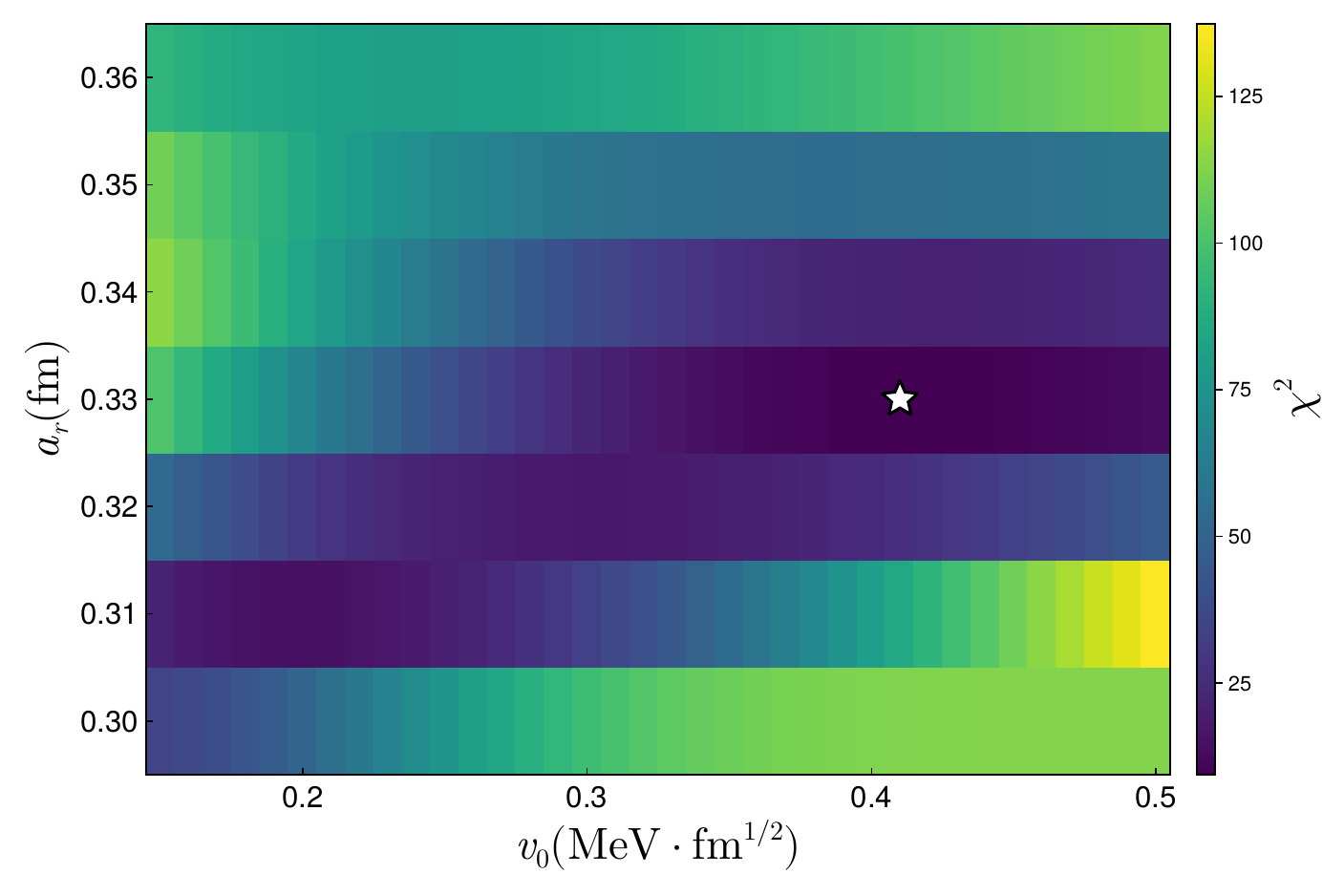}
      \caption{
      The $\chi^2$ values for the astrophysical $S^*$-factor for the ${}^{12}\mathrm{C} + {}^{13}\mathrm{C}$ reaction obtained by varying $a_r$ from 0.30 to 0.36 fm and $v_0$ from 0.15 to 0.50 MeV fm$^{1/2}$. The star symbol indicates the minimum point of $\chi^2$ at $a_r$ = 0.33 fm and $v_0 =$ 0.41 MeV fm$^{1/2}$\label{fig:chi2_plt}.}
\end{figure}

For this purpose, we first determine those parameters for the 
$^{12}$C+$^{13}$C system, as it shows a smooth energy dependence. 
Fig.~\ref{fig:chi2_plt} shows the $\chi^2$ values of the $S^*$-factor for this system, 
\begin{eqnarray}\label{eq:chi2_def}
  \chi^2 = \frac{1}{N_{\rm exp}} \sum_{\mathrm{exp}} \qty(\frac{S^*_{\mathrm{th}}(E_{\mathrm{exp}}) - S^*_{\mathrm{exp}}(E_{\mathrm{exp}})}{\Delta S^*_{\mathrm{exp}}(E_{\mathrm{exp}})})^2,
\end{eqnarray}
where 
$S^*_{\mathrm{exp}}$ and $\Delta S^*_{\mathrm{exp}}$ are the experimental $S^*$-factor and 
its uncertainty, respectively, while $S^*_{\mathrm{th}}$ is the theoretical $S^*$-factor 
at the corresponding energy, $E_{\rm exp}$. 
$N_{\rm exp}$ is the number of the experimental data.
From the figure, 
we find 
the parameter set $v_0=0.41$ MeV fm$^{1/2}$ and $a_r=0.33$ fm gives the minimum value of 
$\chi^2$ and we adopt these values in the following calculations. 

For the $^{12}$C+$^{12}$C system, we use the same value of $a_r$ as that for the 
$^{12}$C+$^{13}$C system. On the other hand, we scale the value of $v_0$ 
so that the $S^*$-factors for the $^{12}\mathrm{C} + {}^{12}\mathrm{C}$ system 
match with those for the $^{12}\mathrm{C} + {}^{13}\mathrm{C}$ system 
at the resonance peaks, as has been observed experimentally. 
This can be done as follows. 
We first obtain 
an effective Hamiltonian for the C+C subspace due to 
the coupling to the Mg states 
by 
projecting Eq.~\eqref{eq:total_hamiltonian} onto the C+C subspace, that is, 
\begin{eqnarray}
  \bm{H}^{l,J^\pi}_{\text{C+C, eff}} = \bm{H}^{l, J^\pi}_{\text{C+C}} + \bm{V} \frac{1}{E_{\rm c.m.} - \bm{H}^{l,J^\pi}_{\text{Mg}}}\bm{V}^T.
\end{eqnarray}\label{eq:effective_hamiltonian}
The imaginary part of this effective Hamiltonian is obtained from the second term as 
\begin{eqnarray}
  \text{Im}\qty(\bm{H}^{l,J^\pi}_{\text{C+C, eff}})_{i, j} = \text{Im} \sum_\mu \frac{v_0^2 \Delta r^{-1} \delta_{i, i_e} \delta_{j, i_e} }{E_{\rm c.m.} - \mathcal{E}_\mu^{J^\pi} + i \Gamma^{J^\pi}(\mathcal{E}_\mu^{J^\pi})/2 } \label{eq:imaginary_pot} .
\end{eqnarray}
The absorption cross section can then be written as
\begin{eqnarray}
  \sigma_{\text{fus}} \propto - w_J \text{Im}\qty(\bm{H}^{l,J^\pi}_{\text{C+C, eff}})_{i_e, i_e} \abs{u(r_{i_e})}^2,
  \label{eq:fusion_cross_imaginary}
\end{eqnarray}
where $w_J$ denotes the spin statistical factor given by
\begin{eqnarray}
  w_J =
    \begin{cases}
    2(2J+1) & \text{for } {}^{12}\mathrm{C}+{}^{12}\mathrm{C},\\[4pt]
    \dfrac{2J+1}{2} & \text{for } {}^{12}\mathrm{C}+{}^{13}\mathrm{C}.
    \end{cases}
\end{eqnarray}
Note that, for the $^{12}\mathrm{C}+{}^{12}\mathrm{C}$ system, an additional factor of 2 is included due to the symmetrization of the initial channel, while 
for the $^{12}\mathrm{C}+{}^{13}\mathrm{C}$ system, a 
factor of $1/2$ is instead included to account for the spin degree of freedom in 
the initial channel.
The ratio of $v_0$ for $^{12}\mathrm{C} + {}^{12}\mathrm{C}$ to that for 
$^{12}\mathrm{C} + {}^{13}\mathrm{C}$
is determined so that Eq.~\eqref{eq:fusion_cross_imaginary} takes the same value 
for both the systems at the resonance energies. 

In the case of $^{12}\mathrm{C}+{}^{12}\mathrm{C}$, that is, for an isolated resonance,
the left hand side of Eq.~\eqref{eq:imaginary_pot} is evaluated as, 
\begin{eqnarray}
  \text{Im}\qty(\bm{H}^{l,J^\pi}_{\text{C+C, eff}})_{i_e, i_e} \propto - \frac{2 v_0^2}{\Gamma^{J^{\pi}}({\cal E}^{J^{\pi}}_\mu)}.
\end{eqnarray}
On the other hand, in the case of $^{12}\mathrm{C}+{}^{13}\mathrm{C}$, that is, for overlapping resonances, it is evaluated as, 
\begin{eqnarray}
  \text{Im}\qty(\bm{H}^{l,J^\pi}_{\text{C+C, eff}})_{i_e, i_e} \propto - \pi v_0^2 \rho({\cal E}^{J^{\pi}}_\mu), 
\end{eqnarray}
where $\rho$ is the level density for $^{25}\mathrm{Mg}$.
Therefore, by requiring that the ratio of the cross sections is unity, one finds
\begin{eqnarray}\label{eq:strength_ratio_equation}
  \frac{(v_0^{12+12})^2}{(v_0^{12+13})^2} = \frac{\pi}{2}\frac{\ev{ w_{J^{25}}\rho^{25}(E_C)} }{\ev{w_{J^{24}}/ \Gamma^{24}(E_C)}} .
\end{eqnarray}
with an assumption that $|u(r_{i_e})|^2$ is similar between the two systems.
Here, $\ev{w_{J^{24}}/\Gamma^{24}(E_C)}$ denotes the energy- and spin-averaged product of the inverse decay width and the spin statistical factor for~$^{24}\mathrm{Mg}$, while $\ev{w_{J^{25}}\rho^{25}(E_C)}$ denotes the energy- and spin-averaged product of the level density and the spin statistical factor for~$^{25}\mathrm{Mg}$.

In the energy range of $2.0~\mathrm{MeV} < E_{\mathrm{c.m.}} < 5.5~\mathrm{MeV}$, the dominant partial waves for the $^{12}\mathrm{C}+{}^{12}\mathrm{C}$ fusion reaction are found to be $l=0$ and $2$, whereas those for the $^{12}\mathrm{C}+{}^{13}\mathrm{C}$ fusion reaction are $l=0$, $1$, and $2$. Therefore, 
we take the averages in Eq.~(\ref{eq:strength_ratio_equation}) with these partial waves. 
More specifically, we take the averages of the decay widths of $^{24}\mathrm{Mg}$ for $J=0$ and $2$, as well as that of the level densities of $^{25}\mathrm{Mg}$ for $J=1/2$, $3/2$, and $5/2$, 
in the energy range of $2.0~\mathrm{MeV} < E_{\mathrm{c.m.}} < 5.5~\mathrm{MeV}$. 
From the shell-model calculation for $\rho^{25}$ and the width $\Gamma^{24}$ 
estimated statistically with Eq. ~\eqref{eq:decay_width}, we obtain
\begin{eqnarray}
  \ev{w_{J^{24}}/\Gamma^{24}(E_C)} &\simeq 24.1 \, \mathrm{MeV}^{-1}  ,\\
  \ev{ w_{J^{25}}\rho^{25}(E_C)} &\simeq 39.8 \, \mathrm{MeV}^{-1}, 
\end{eqnarray}
from which the value of $v_0$ for the $^{12}$C+$^{12}$C reaction is determined to be 
$v_0 = 0.66~\mathrm{MeV}\,\mathrm{fm}^{1/2}$. 

\subsection{Fusion cross sections for the $^{12}$C+$^{12,13}$C reactions}

\begin{figure}[tbp]
        \includegraphics[width=\linewidth]{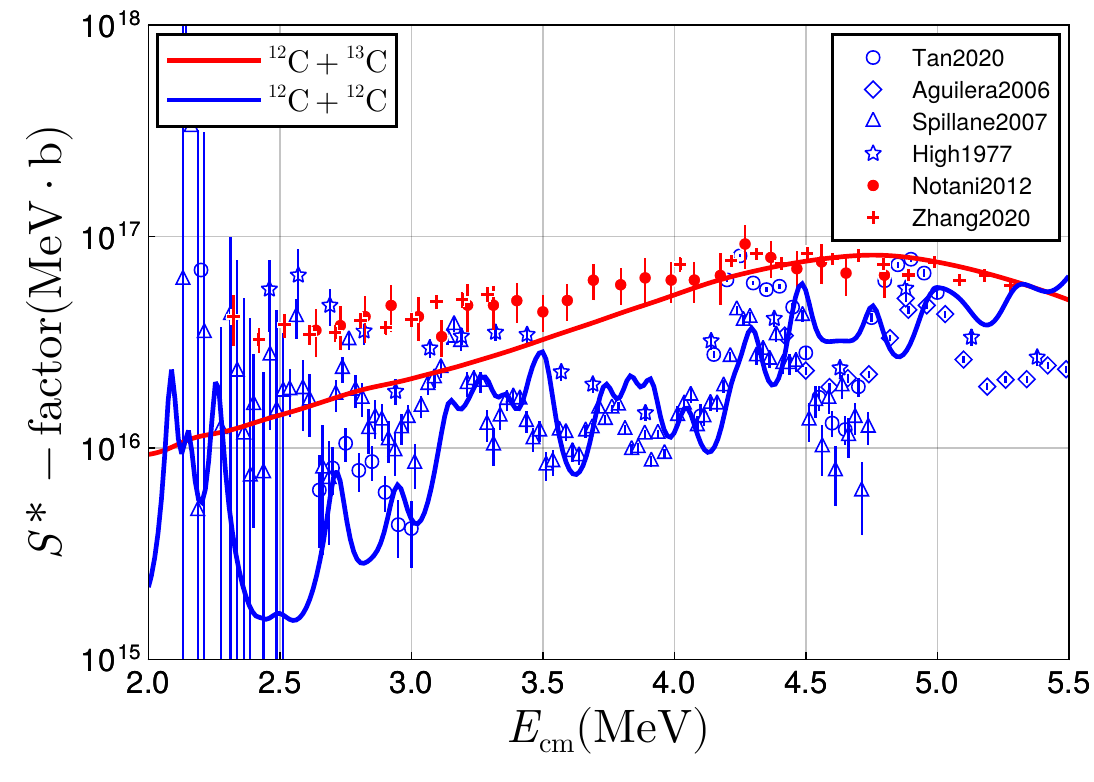}
        \caption{Best-fit results for the $S^*$-factors of  $^{12}\mathrm{C} + {}^{12}\mathrm{C}$ (the blue solid curve) and $^{12}\mathrm{C} + {}^{13}\mathrm{C}$ (the red solid curve) fusion reactions. The experimental data for $^{12}\mathrm{C}+{}^{12}\mathrm{C}$ are taken from Refs.~\cite{ PhysRevC.73.064601,PhysRevLett.124.192702,HIGH1977181, PhysRevLett.98.122501}, and those for $^{12}\mathrm{C}+{}^{13}\mathrm{C}$ are taken from Refs.~\cite{ZHANG2020135170,PhysRevC.85.014607}.}
        \label{fig:m3y_rep_final_result}
\end{figure}

\begin{figure*}[tb]
    \begin{minipage}{0.49\linewidth}
        \includegraphics[width=\linewidth]{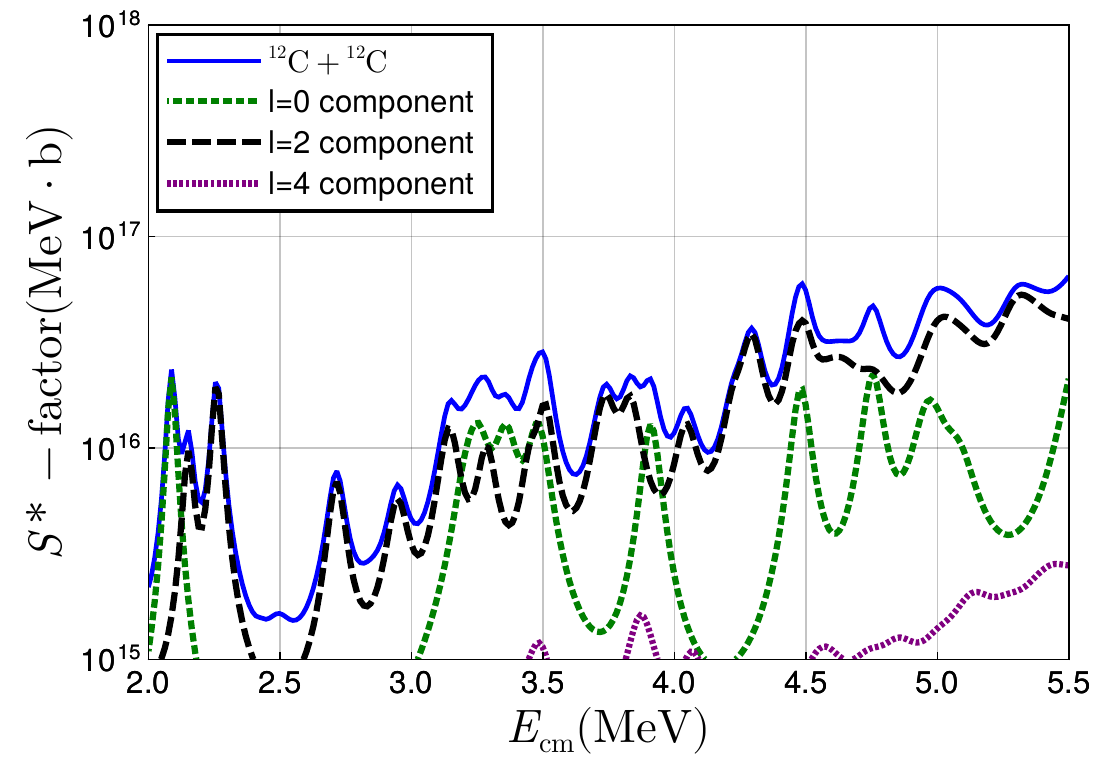}
    \end{minipage}
    \hfill
    \begin{minipage}{0.49\linewidth}
        \includegraphics[width=\linewidth]{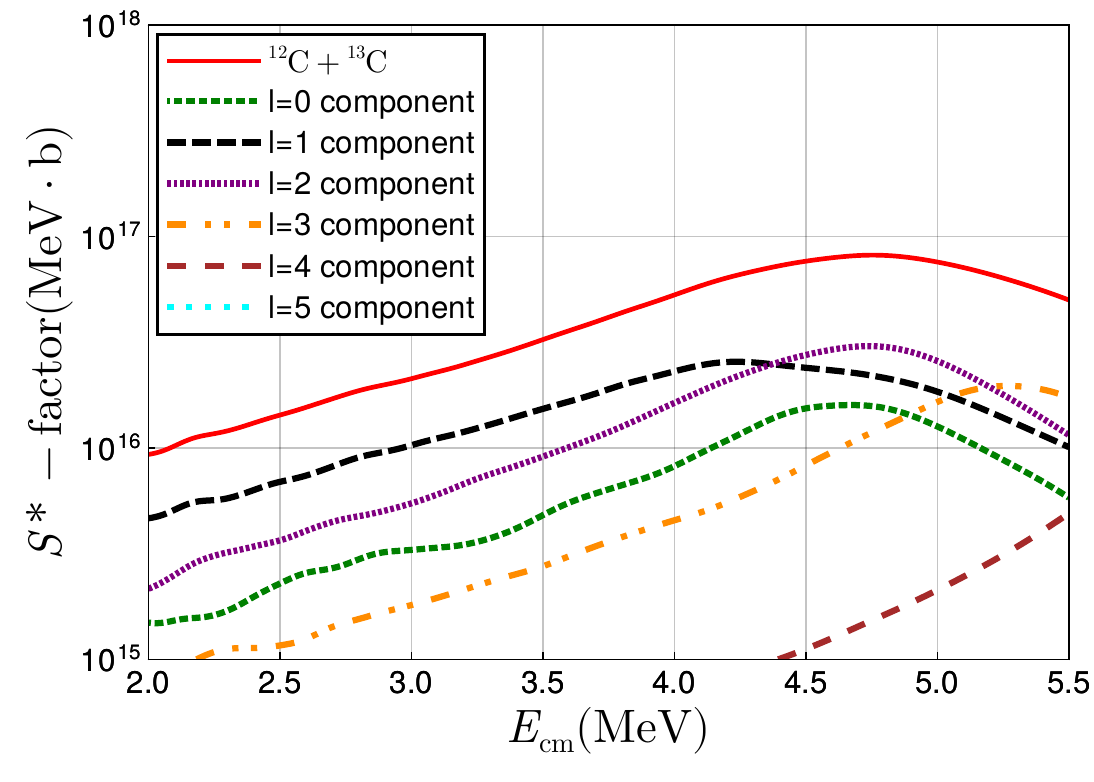}
    \end{minipage}
  \caption{The contributions from each partial waves, as shown in the insets of the figures. The left panel shows the astrophysical $S^*$-factors for the $^{12}\mathrm{C}+{}^{12}\mathrm{C}$ system, while 
  the right panel shows those for the $^{12}\mathrm{C}+{}^{13}\mathrm{C}$ system. \label{fig:each_partial_wave}}
\end{figure*}

With the parameters $v_0$ and $a_r$ determined in the previous subsection, 
we now calculate fusion cross sections. 
The results for the two systems are plotted in the same figure, Fig.~\ref{fig:m3y_rep_final_result}.
From the figure, one finds that the calculated $S^*$-factor for the~$^{12}\mathrm{C}+{}^{12}\mathrm{C}$ system 
reproduces the prominent resonance structures with comparable magnitudes 
to the experimental data 
in the energy region where the experimental data exist. 
The $S^*$-factors for the ~$^{12}\mathrm{C}+{}^{13}\mathrm{C}$ system are also reproduced reasonably well at energies above around 3.5 MeV. In particular, a smooth energy dependence is well reproduced. 
Our calculation underestimates the $S^*$-factors at lower energies. 
This may be because we neglected the internal excitations of the carbon nuclei in the C+C channel, 
as well as the elastic one-neutron transfer for the $^{12}$C+$^{13}$C system.

Fig.~\ref{fig:each_partial_wave} shows the contribution of each partial wave component.
In the low-energy region, the number of levels in $^{24}\mathrm{Mg}$ is small 
 as shown in Fig.~\ref{fig:accumulated_num_levels}, 
and at the same time, the decay widths also become small 
as shown in Fig.~\ref{fig:decay_width}. 
Because of these effects, 
the interval of the off-resonance region, in which no level of $^{24}$Mg exists 
at the energy corresponding to a given center-of-mass energy, tends to be wide. 
As a result, the $S^*$ factor takes considerably different values (by almost one order of magnitude) depending upon whether a resonance is present or not at energies close to an incident energy.

\subsection{The effects of decay width}
\begin{figure}[tb]
 \centering
        \includegraphics[width=\linewidth]{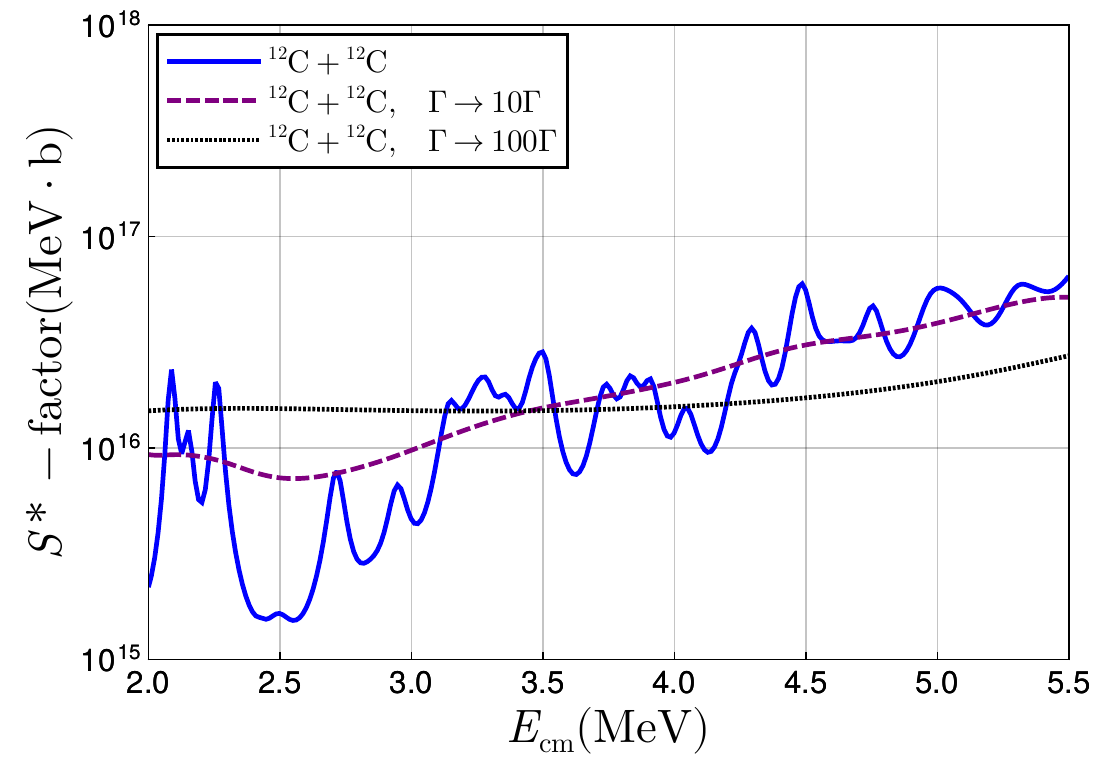}
        \caption{The dependence of the $S^*$-factors for $^{12}\mathrm{C}+^{12}\mathrm{C}$ fusion 
        on the decay width of the compound nucleus, $^{24}$Mg. The blue curve shows the original 
        result shown in Fig. \ref{fig:m3y_rep_final_result}. 
        The purple dashed and the black dotted curves show the results 
        with larger decay widths, by a factor of 10 and 100, respectively. 
The other parameters are taken to be the same as the original calculation.}
        \label{fig:12c_12c_decay_width_change}
\end{figure}

\begin{figure}[tb]
    \centering
        \includegraphics[width=\linewidth]{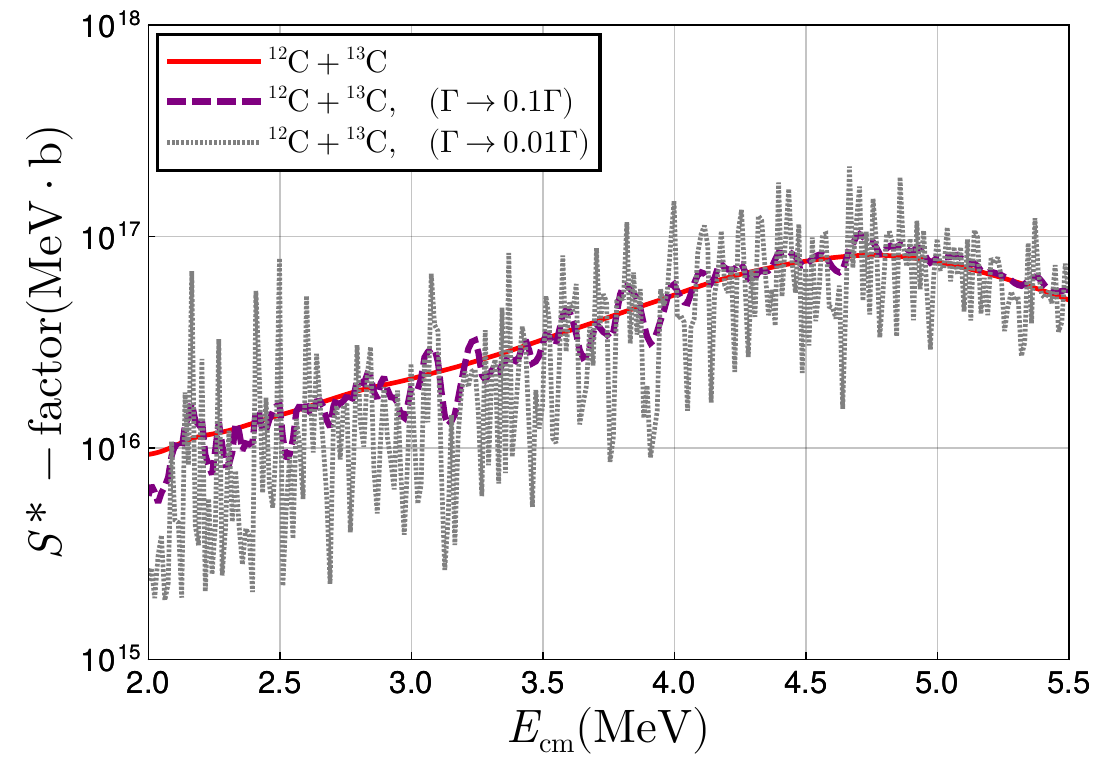}
        \caption{
        Same as Fig. \ref{fig:12c_12c_decay_width_change} but for the $^{12}\mathrm{C}+^{13}\mathrm{C}$ 
        system. The red curve shows the original result, while the purple dashed and the gray dotted curves are obtained by reducing the decay width of $^{25}$Mg by factors of 0.1 and 0.01, respectively.  }
        \label{fig:12c_13c_decay_width_change}
\end{figure}

For compound nucleus resonances,
the relation between the average decay width $\Gamma$ and the average level spacing $D$ is important \cite{PhysRev.157.907}.
In the case of $^{12}\mathrm{C} +{}^{12}\mathrm{C}$ fusion reaction, the level density of $^{24}\mathrm{Mg}$ is low and the resonance peaks are separated from each other\cite{PhysRevLett.110.072701}. 
To clarify the role of the decay width in the energy dependence of $S^*$-factors, 
in this subsection we repeat the calculations by varying the decay width while keeping the 
other parameters unchanged. 
That is, we make the decay width of $^{24}$Mg larger and check 
whether the $S^*$-factors show the smooth energy dependence as a result of 
the overlapping resonance peaks. 

Fig.~\ref{fig:12c_12c_decay_width_change} shows the $S^*$-factors for the 
 $^{12}\mathrm{C} +^{12}\mathrm{C}$ system with larger values of the decay width, $\Gamma$. 
 The purple dashed and the black dotted lines are obtained by multiplying the decay width by a factor of 
 10 and 100, respectively, while the blue solid curve is the original results shown in Fig. 
 \ref{fig:m3y_rep_final_result}. 
As one can see in the figure, as the decay width increases, the resonance structure disappears and 
the cross sections, and thus the $S^*$-factors, show a smooth energy dependence.

In the case of $^{12}\mathrm{C} + {}^{13}\mathrm{C}$, the $S^*$-factors show a smooth energy dependence with the original decay width. 
Let us then make the decay width smaller and check whether the resonance peaks become 
isolated from each other. The results are shown in Fig. \ref{fig:12c_13c_decay_width_change}. 
As is expected, the $S^*$-factors show prominent resonance structures as the decay width decreases. 

These studies support the interpretation that the smooth excitation function in $^{12}\mathrm{C}+{}^{13}\mathrm{C}$ is associated with overlapping compound resonances, whereas the pronounced resonance structures in $^{12}\mathrm{C}+{}^{12}\mathrm{C}$ are due to relatively isolated resonances.
Here we changed only the decay widths while the level spacings were unmodified. The situation of 
the overlapping compound resonances can be realized also by decreasing the level spacing with keeping 
the decay width. In reality, both of the decay width and the level spacing 
are involved in the different behaviors 
between the $^{12}\mathrm{C}+{}^{12}\mathrm{C}$ and the $^{12}\mathrm{C}+{}^{13}\mathrm{C}$ systems.

\subsection{The effects of the underestimation of the level density in $^{24}\mathrm{Mg}$}

\begin{figure}[tb]
\centering
        \includegraphics[width=\linewidth]{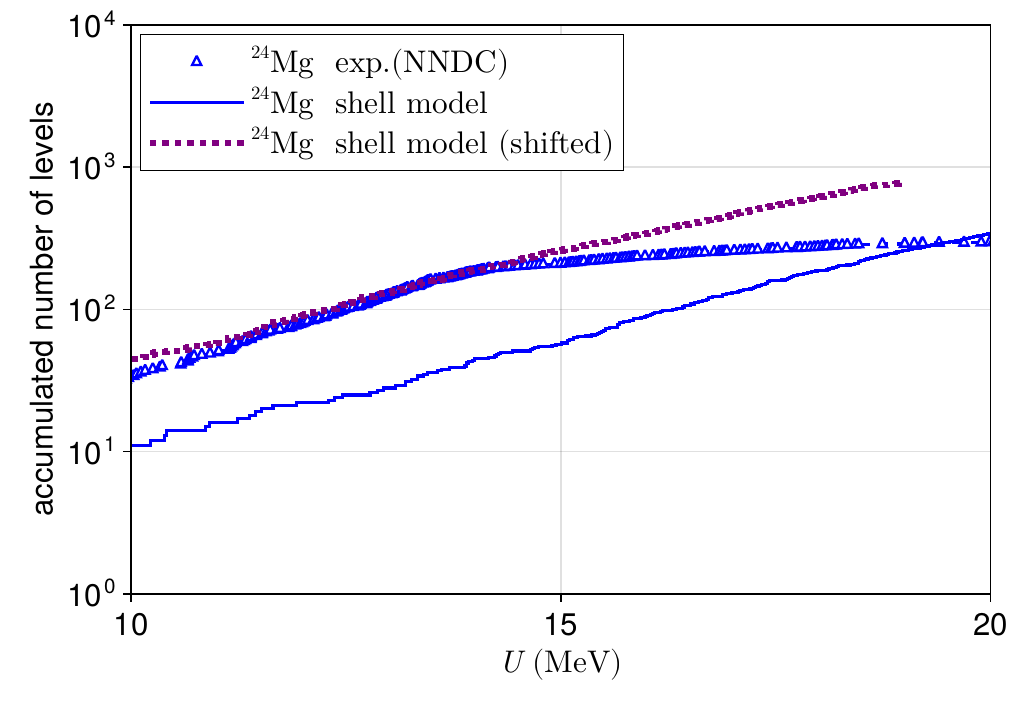}
        \caption{Same as Fig. \ref{fig:accumulated_num_levels} for $^{24}$Mg, but shifted by 4 MeV to reproduce the experimental accumulated number of levels (the purple dotted curve).
}
        \label{fig:shifted_shell_model_levels}
\end{figure}

\begin{figure}[tb]
        \includegraphics[width=\linewidth]{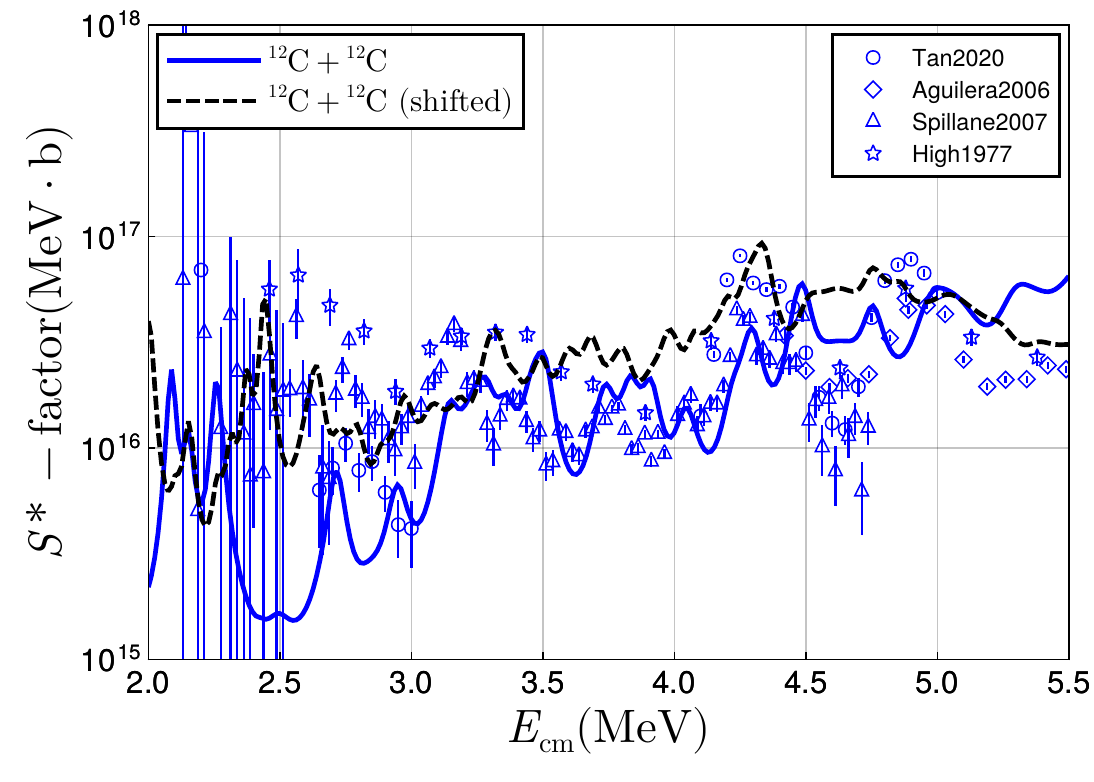}
        \caption{
Same as Fig. \ref{fig:m3y_rep_final_result} for $^{12}$C+$^{12}$C, but with the $S^*$-factors obtained with the shifted energy spectrum shown in Fig. \ref{fig:shifted_shell_model_levels} (the black dashed curve). 
        }
        \label{fig:shifted_sfactor}
\end{figure}

As shown in Fig. \ref{fig:accumulated_num_levels}, 
the \verb+sdpf-mu+ interaction 
significantly underestimates the level density of the $^{24}\mathrm{Mg}$. 
How would this affect the energy dependence 
of the $S^*$-factors for the $^{12}$C+$^{12}$C reaction? 
To check this, we repeat the calculation by shifting the calculated level 
spectrum of \verb+KSHELL+, that is, by modifying ${\cal E}_\mu$ in Eq. (\ref{eq:mg_hamiltonian})
by ${\cal E}_\mu\to {\cal E}_\mu-\Delta{\cal E}$ with $\Delta{\cal E}=4.0$ MeV, as shown in 
Fig~\ref{fig:shifted_shell_model_levels}. 
The resultant $S^*$-factors are shown in Fig.~\ref{fig:shifted_sfactor}.
 As one can see, the oscillatory behavior is somewhat suppressed but the resonance behavior does not disappear, especially around 2.5 MeV. On the other hand, at higher energies above 4.5 MeV, the calculated $S^*$ factors with the shifted spectrum show a smooth energy dependence due to 
the high level density. 
 However, the experimental data still exhibit several resonance structures in this higher energy region. This may suggest that the resonance structures observed experimentally above 4.5 MeV do not originate from the individual dense compound states of ${}^{24}\text{Mg}$, 
even though one may need to investigate the channel coupling effects, which are ignored in this paper.  

\section{summary\label{sec:summary}}

In this study, we have calculated the fusion cross sections of the $^{12}\mathrm{C}+{}^{12}\mathrm{C}$ and $^{12}\mathrm{C}+{}^{13}\mathrm{C}$ reactions in the astrophysical energy region within a unified model that explicitly incorporates couplings of the reaction channel to compound nucleus states. In this model, the compound-nucleus states of the Mg isotopes have been described microscopically 
by shell-model calculations, while their decay widths have been estimated based on the statistical 
model. In this way, we have treated the formation of a compound nucleus with a physically more grounded 
model than the phenomenological optical potential model. 
 We have succeeded in reproducing in a unified manner the experimentally observed behaviors of the two fusion reactions, that is, the smooth energy dependence of fusion cross sections for the 
 $^{12}\mathrm{C}+{}^{13}\mathrm{C}$ system and the pronounced resonance behaviors in the 
 $^{12}\mathrm{C}+{}^{12}\mathrm{C}$ system. 
 The present study 
 indicates that the difference between the two fusion reactions can be understood in terms of the properties of the compound nucleus states. Namely, the level density as well as the decay width of the compound nucleus play essential roles in determining the different resonance behaviors of the two systems. 

The present model can be extended in several ways. 
Firstly, one can replace 
the decay widths by a Hamiltonian that explicitly describes the relative motion in the 
final channels, such as $\alpha + \mathrm{Ne}$. 
This extension would make it possible to incorporate interference effects that are not taken into account in the present framework, namely those arising from transitions to the same final state through different excited states of the compound nucleus. 
Furthermore, one could release the restriction that the coupling between the entrance channel 
and the compound nucleus states takes place at a single radial position with an 
equal strength for all the compound nucleus states. This could be incorporated with 
the resonating group method (RGM) as in Ref. \cite{PhysRevC.79.044606}. 
In addition, 
by including the channel coupling effects in the entrance channel, 
a more quantitative description of the fusion cross sections will be achieved.
We will report these extensions in a separate publication. 

\section*{Acknowledgments}

We thank George F. Bertsch and Alex B. Brown for useful discussions at the early stage of this work.  
We also thank Xiaodong Tang for useful discussions.
This work was supported in part by JSPS KAKENHI Grant Numbers JP23K03414 and JP26K17162, and by JST SPRING, Grant Number JPMJSP2110. 
The shell-model calculations were carried out at Yukawa Institute Computer Facility.
\bibliography{paper}

\end{document}